\documentclass[preprint,eqsecnum,aps,nofootinbib]{revtex4}
\usepackage{amsfonts,amsmath,amssymb,amsthm}
\usepackage{latexsym}
\usepackage{bbm,bm}
\usepackage{graphicx}
\usepackage{tikz} 
\usepackage{multirow}

\newcommand{\ket}[1]{\lvert #1 \rangle}
\newcommand{\bra}[1]{\langle #1 \lvert}
\newcommand{\beq}{\begin{equation}}
\newcommand{\eeq}{\end{equation}}
\newcommand{\beqs}{\begin{eqnarray}}
\newcommand{\eeqs}{\end{eqnarray}}

\begin{document}

\title{Euclidean time method in Generalized Eigenvalue Equation }

\author{Mi-Ra Hwang$^1$, Eylee Jung$^1$, MuSeong Kim$^2$ and DaeKil Park$^{1,3}$\footnote{corresponding author, dkpark@kyungnam.ac.kr} }

\affiliation{$^1$Department of Electronic Engineering, Kyungnam University, Changwon,
                 631-701, Korea    \\
                $^2$Pharos iBio Co., Ltd. 
                Head Office: \#1408, 38, Heungan-daero 427beon-gil, Dongan-gu, Anyang, 14059, Korea \\
                $^3$Department of Physics, Kyungnam University, Changwon,
                  631-701, Korea }

\begin{abstract}
We develop the Euclidean time method of the variational quantum eigensolver for solving the generalized eigenvalue equation $A \ket{\phi_n} = \lambda_n B \ket{\phi_n}$, 
where $A$ and $B$ are hermitian operators, and $\ket{\phi_n}$ and $\lambda_n$ are called the eigenvector and the corresponding eigenvalue of this equation respectively.
For the purpose we modify the usual Euclidean time formalism, which was developed for solving the time-independent Schr\"{o}dinger equation.
We apply our formalism to three numerical examples for test. It is shown that our formalism works very well in all numerical examples. 
We also apply our formalism to the hydrogen atom and compute the electric polarizability. It turns out that our result is slightly less than that of the perturbation method. 

\end{abstract}

\maketitle

\section{Introduction}
After Feynman's suggestion on quantum computer\cite{feynman82,feynman86} few decades ago, the hardwares and the algorithms are rapidly developed recently. 
The quantum computers with few hundred qubits were constructed in several companies such as IBM and Google. Also, some quantum algorithms have been presented 
such as factoring\cite{text,shor94}, database searching\cite{text,grover96}, and matrix inversion\cite{hhl09}. However, it seems to be far away to construct the large-scale, fault-tolerant 
universal quantum computer. 

In spite of this fact, the current-stage quantum computers have their own merits when the classical computers are simultaneously used. 
In this reason, the hybrid quantum-classical algorithms play important role recently on 
noisy intermediate-scale quantum (NISQ) era. The representatives of the hybrid algorithm are the quantum approximate optimization algorithm (QAOA)\cite{farhi14}
and variational quantum eigensolver (VQE)\cite{peruzzo13}. QAOA has been used to find approximate solutions of classical Ising models\cite{ising17} and 
clustering problems formulated as MaxCut\cite{max_cut}.
VQE was first applied in Ref.\cite{peruzzo13} to compute the ground state molecular energy for helium hydride ion $HeH^+$.
The hybrid quantum-classical algorithms have been used in finding the energy spectra\cite{malley15,higgott18,endo18,vogt20}, simulating the Schr\"{o}dinger equations\cite{ying16,mahdian20,endo18-2},
and quantum machine learning\cite{benedetti19,wang20,keren18}. They were also applied to black hole physics\cite{bh21}, high-energy physics\cite{hep22}, and cosmology\cite{cosmos22}.

VQE is a variational algorithm designed to find the ground state of a system governed by a Hamiltonian $H$. Let $\ket{\phi}$ be an initial state that is easy to prepare. By applying a unitary 
operator $U({\bm \theta})$ we prepare the parameter-dependent quantum state:
\begin{equation}
\label{initial-1}
\ket{\psi ({\bm \theta})} = U({\bm \theta}) \ket{\phi}.
\end{equation}
Then, the expectation value of the Hamiltonian can be written as 
\begin{equation}
\label{expectation-1}
E({\bm \theta}) = \bra{\psi ({\bm \theta})} H \ket {\psi ({\bm \theta})}.
\end{equation} 
If $U({\bm \theta})$ is selected appropriately, the ground state energy $E_0$ can be computed by minimizing $E({\bm \theta})$:
\begin{equation}
\label{variation-1}
E_0 = E({\bm \theta}_{min}) = \min_{\bm \theta} E({\bm \theta}).
\end{equation}
Furthermore, the ground state $\ket{\psi_0}$ can be derived as $\ket{\psi_0} = \ket{\psi ({\bm \theta}_{min})}$. This is a whole story of the variational method in quantum mechanics. 
In VQE the quantum computer computes expectation value in Eq. (\ref{expectation-1}) while the minimization of $E({\bm \theta})$ is carried out in the classical computer. 

In many papers the classical computer uses the classical optimizers such as Nelder-Mead for the minimization of  $E({\bm \theta})$. However, the classical optimizers can 
yield an incorrect answer if $E({\bm \theta})$ has local minima. Even though there are several methods\cite{lo-minima} to escape the local minima problem, we think that the most physically 
appealing method is a Euclidean time method introduced in Ref.\cite{euclidean-1}, because the Euclidean time $\tau = i t$ is frequently used in the path-integral quantum mechanics\cite{feynman,kleinert}.
For example, let us consider the one-dimensional simple harmonic oscillator system. Then, the Euclidean propagator is 
\begin{equation}
\label{sho-1}
G[x_b, x_a: \tau] = \sqrt{\frac{m \omega}{2 \pi \hbar \sinh \omega \tau}} \exp\left[ - \frac{m \omega}{2 \hbar \sinh \omega \tau} 
\left\{ (x_a^2 + x_b^2) \cosh \omega \tau - 2 x_a x_b \right\} \right].
\end{equation}
If we take $\tau \rightarrow \infty$ limit, the propagator becomes 
\begin{equation}
\label{sho-2}
\lim_{\tau \rightarrow \infty} G[x_b, x_a: \tau] \sim \phi_0^* (x_b) \phi_0 (x_a) e^{-\frac{i}{\hbar} E_0 \tau}
\end{equation}
where 
\begin{equation}
\label{sho-3}
\phi_0 (x) = \left(\frac{m \omega}{\pi \hbar} \right)^{1/4} e^{-\frac{m \omega}{2 \hbar} x^2}  \hspace{1.0cm} E_0 = \frac{1}{2} \hbar \omega.
\end{equation}
These are exact eigenfunction and eigenvalue for the ground state of the system. Thus, the VQE with the Euclidean time\cite{euclidean-1} naturally yields ground state energy and the corresponding eigenvector at  large $\tau$ limit. 
This technique was used to discover Hamiltonian spectra\cite{spectra} and  is extended to the mixed state scenario\cite{mixed}.

In this paper we want to apply the Euclidean time method of VQE to the generalized eigenvalue equation(GEE)
\begin{equation}
\label{gevalue-1}
A \ket{\phi_n} = \lambda_n B \ket{\phi_n}  \hspace{1.0cm} (n = 0, 1, \cdots)
\end{equation}
where $A$ and $B$ are hermitian operators, and $\lambda_n$ is a generalized eigenvalue. 
The GEE was used in Ref.\cite{gee1969} to compute the electric polarizability in the hydrogen atom. 
GEE problems also arise in the quantum chemistry\cite{fordgee} and fluid mechanics\cite{fluidgee}.
In order to solve Eq. (\ref{gevalue-1}) in quantum computer the slightly variant of the quantum phase estimation (QPE) was suggested in Ref.\cite{qpegee}.
However, the QPE technique generally requires long coherence time and hence, is not suitable for the NISQ devices. 
In order to overcome the difficulty, the authors in Ref.\cite{lslf22} used the quantum gradient descent algorithm and solve the numerical example.

The paper is organized as follows. In section II we present a formalism, which shows how to apply the Euclidean time method to the generalized eigenvalue problem (GEP).
In section III we solve the numerical example of the GEP when $B$ is regular operator. In section IV we consider another numerical problem when $B$ is singular operator\footnote{ If $B$ is regular, the GEE (\ref{gevalue-1}) can be converted into the usual eigenvalue equation 
$\left(B^{-1} A \right)  \ket{\phi_n} = \lambda_n  \ket{\phi_n}$ by incorporating the matrix inversion algorithm \cite{hhl09} in principle. If, however, $B$ is singular, such a conversion is impossible because $B^{-1}$ does not exist.}. It is shown that 
the Euclidean time technique introduced in this paper works very well when $B$ is regular or singular. In section V we introduce another numerical example, where $A$ and $B$ are $8 \times 8$ matrices. It turns out that the eigenvalues converges very slowly with respect to the Euclidean time $\tau$ 
compared to the previous numerical examples.
In section VI we review Ref. \cite{gee1969}, where the electric polarizability of the hydrogen atom is calculated perturbatively  by applying the GEE. In section VII we explore the same atomic physics issue by applying the Euclidean time method. It turns out that the result of this section 
is slightly less than that of the perturbation method. In section VIII a brief conclusion is given. In appendix A we summarize the calculation of section VII as a Table II.


\section{Formalism}
Let us consider the GEE of Eq. (\ref{gevalue-1}). 
Due to the matrix $B$ the orthogonality of the normalized eigenvectors is expressed as
\begin{equation}
\label{gevalue-2}
\bra{\phi_m} B \ket{\phi_n} = \delta_{mn}.
\end{equation}
In the following we will call the condition (\ref{gevalue-2}) by $B$-orthogonality.


We start with a generalized Euclidean time-dependent Schr\"{o}dinger-like equation
\begin{equation}
\label{imaginary-1}
\frac{\partial}{\partial \tau} \ket{\psi(\tau)} = - (A - \lambda B)  \ket{\psi(\tau)}
\end{equation}
where $\tau = i t$ is an Euclidean time. If $\ket{\psi(\tau)}$ is an eigenvector of Eq. (\ref{gevalue-1}), the eigenvalue $\lambda$ in Eq. (\ref{imaginary-1})
can be written as 
\begin{equation}
\label{imaginary-2}
\lambda \rightarrow F(\tau) = \frac{\bra{\psi(\tau)} A \ket{\psi(\tau)}} {\bra{\psi(\tau)} B \ket{\psi(\tau)}}.
\end{equation}
Thus, the Euclidean time evolution of $\ket{\psi(\tau)}$ is governed by 
\begin{equation}
\label{imaginary-3}
\frac{\partial}{\partial \tau} \ket{\psi(\tau)}= - (A - F(\tau) B) \ket{\psi(\tau)}.
\end{equation}
As usual Euclidean quantum mechanics, $\ket{\psi(\tau)}$ should approach to the ground state of Eq. (\ref{gevalue-1}) in the $\tau \rightarrow \infty$ limit.
Thus, we want to solve Eq. (\ref{imaginary-3}) by applying the hybrid quantum-classical algorithm.

In order to solve Eq. (\ref{imaginary-3}) numerically, we assume $\ket{\psi(\tau)}$ as 
\begin{eqnarray}
\label{vqa-1}
&& \hspace{1.0cm} \ket{\psi(\tau)} = V ({\bm \theta}) \ket{\bar{0}}    \\    \nonumber
&& V ({\bm \theta}) = U_N (\theta_N) \cdots U_k (\theta_k) \cdots U_1 (\theta_1)
\end{eqnarray}
where $U_k$ is an unitary operator and $\theta_k$ is dependent on $\tau$. Using 
\begin{equation}
\label{vqa-2}
\frac{\partial}{\partial \tau} \ket{\psi(\tau)} = \sum_{i=1}^N \dot{\theta}_i \frac{\partial}{\partial \theta_i} \ket{\psi(\tau)} ,
\end{equation}
one can show directly 
\begin{eqnarray}
\label{vqa-3}
&& \hspace{3.0cm}  \left| \left( \frac{\partial}{\partial \tau} + A - F B \right) \ket{\psi(\tau)} \right|^2               \\   \nonumber
&& = \sum_{i,j=1}^N \dot{\theta}_i \dot{\theta}_j \left( \frac{\partial}{\partial \theta_j} \bra{\psi(\tau)} \right) \left( \frac{\partial}{\partial \theta_i} \ket{\psi(\tau)} \right) 
+ \bra{\psi(\tau)} (A - F B)^2 \ket{\psi(\tau)}                                                                                \\   \nonumber
&&\hspace{.5cm}+  \sum_{i=1}^N \dot{\theta}_i \bra{\psi(\tau)} (A - F B) \left( \frac{\partial}{\partial \theta_i} \ket{\psi(\tau)} \right) 
+ \sum_{i=1}^N \dot{\theta}_i \left( \frac{\partial}{\partial \theta_i} \bra{\psi(\tau)} \right) (A - F B)  \ket{\psi(\tau)}.
\end{eqnarray}
Applying the McLachlan's variational principle
\begin{equation}
\label{vqa-4}
\frac{\partial}{\partial \dot{\theta}_j} \left| \left( \frac{\partial}{\partial \tau} + A - F B \right) \ket{\psi(\tau)} \right| = 0,
\end{equation}
one can derive the first-order coupled differential equation of the parameter
\begin{equation}
\label{vqa-5}
\sum_{j=1}^N \Gamma_{ij} \dot{\theta}_j = C_i
\end{equation}
where 
\begin{eqnarray}
\label{vqa-6}
&&\Gamma_{ij} = \mbox{Re} \left[ \left( \frac{\partial}{\partial \theta_i} \bra{\psi(\tau)} \right) \left( \frac{\partial}{\partial \theta_j} \ket{\psi(\tau)} \right)  \right]   \\    \nonumber
&& C_i = - \mbox{Re} \left[ \left( \frac{\partial}{\partial \theta_i} \bra{\psi(\tau)} \right) (A - F B)  \ket{\psi(\tau)} \right].
\end{eqnarray}

The hybrid quantum-classical algorithm we adopt in this paper is as following. We solve the differential equation (\ref{vqa-5}) in the classical computer by making use of the Euler method
\begin{equation}
\label{euler-1}
{\bm \theta} (\tau + \delta \tau) \approx {\bm \theta} (\tau) + \Gamma^{-1} (\tau) {\bm C} (\tau) \delta \tau.
\end{equation}
The coefficients $\Gamma_{ij} (\tau)$ and $C_i (\tau)$ as well as $F(\tau)$ in Eq. (\ref{imaginary-2}) will be computed via the suitable quantum algorithms. 

\begin{figure}[ht!]
\begin{center}
\includegraphics[height=4.0cm]{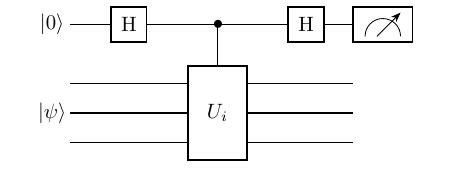} 
\caption[fig1]{(Color online) If $U_i$ is unitary as well as hermitian, $\bra{\psi} U_i \ket{\psi}$ can be computed by $P(0) - P(1)$ in this circuit, where $P(j)$ is a probability for outcome $j$.                       }
\end{center}
\end{figure}

First, let us briefly comment how to compute $F(\tau)$. We assume that the matrices $A$ and $B$ are $2^m \times 2^m$ hermitian. Then, $A$ and $B$ can be decomposed into the linear combination 
of $m$-tensor product of the Pauli matrices:
\begin{eqnarray}
\label{pauli}
\sigma_0 = I_2 = \left(  \begin{array}{cc} 1 & 0  \\  0 & 1  \end{array}   \right)            \hspace{.5cm}
\sigma_1 = \left(  \begin{array}{cc} 0 & 1  \\  1 & 0  \end{array}   \right)            \hspace{.5cm}
\sigma_2 = \left(  \begin{array}{cc} 0 & -i  \\  i & 0  \end{array}   \right)            \hspace{.5cm}
\sigma_3 = \left(  \begin{array}{cc} 1 & 0  \\  0 & -1  \end{array}   \right).           
\end{eqnarray}
Since Pauli matrices are unitary as well as hermitian, the expectation value of each term, say $U_i$, can be computed by applying Fig. 1. 
In this way it is possible to compute $F(\tau)$ by applying the circuit of Fig. 1 repeatedly.

\begin{figure}[ht!]
\begin{center}
\includegraphics[height=2.2cm]{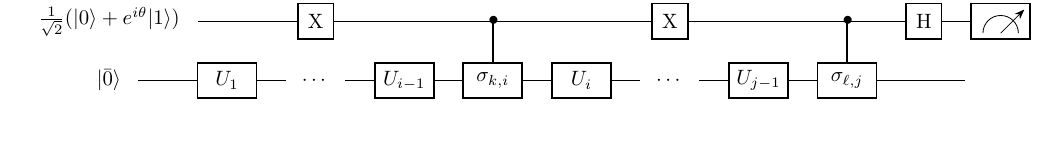}  
\includegraphics[height=2.2cm]{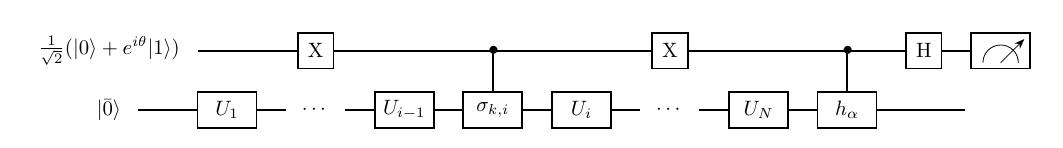}

\caption[fig2]{(Color online) (a) One can compute $\mbox{Re} \left[ e^{i \theta} \bra{\bar{0}} \widetilde{V}_{k,i}^{\dagger} \widetilde{V}_{\ell,j} \ket{\bar{0}} \right]$ by measuring $P(0) - P(1)$ in the first circuit
where $P(j)$ is a probability for outcome $j$. (b) One can compute $\mbox{Re} \left[ e^{\i \theta} \bra{\bar{0}} \widetilde{V}_{k,i}^{\dagger} h_{\alpha} V \ket{\bar{0}} \right]$ by measuring $P(0) - P(1)$ in the second circuit.
 }
\end{center}
\end{figure}

Now, let us explain how to compute $\Gamma_{ij}$ and $C_i$ with quantum circuits. Let us express the derivative of $U_i (\theta_i)$ in a form:
\begin{equation}
\label{vqa-7}
\frac{\partial U_i (\theta_i)}{\partial \theta_i} = \sum_{k=1}^N f_{k,i} U_i (\theta_i) \sigma_{k,i},
\end{equation}
where $f_{k,i}$ is a complex number and $\sigma_{k,i}$ is unitary operator. Then, one can show easily 
\begin{equation}
\label{vqa-8}
\frac{\partial}{\partial \theta_i} \ket{\psi(\tau)} = \sum_k f_{k,i} \widetilde{V}_{k,i} \ket{\bar{0}}
\end{equation}
where 
\begin{equation}
\label{vqa-9}
\widetilde{V}_{k,i} = U_N \cdots U_i \sigma_{k,i} U_{i-1} \cdots U_1.
\end{equation}
Inserting Eq. (\ref{vqa-8}) into Eq. (\ref{vqa-6}), one can show 
\begin{eqnarray}
\label{vqa-10}
&&  \Gamma_{ij} = \mbox{Re} \left[ \sum_{k,\ell = 1}^N f_{k,i}^* f_{\ell,j} \bra{\bar{0}} \widetilde{V}_{k,i}^{\dagger} \widetilde{V}_{\ell,j} \ket{\bar{0}} \right]             \\   \nonumber
&& C_i = - \mbox{Re} \left[ \sum_{k,\alpha} f_{k,i}^* \Lambda_{\alpha} \bra{\bar{0}} \widetilde{V}_{k,i}^{\dagger} h_{\alpha} V \ket{\bar{0}} \right]
\end{eqnarray}
where we used a decomposition 
\begin{equation}
\label{vqa-11}
A - F B = \sum_{\alpha} \Lambda_{\alpha} h_{\alpha}.
\end{equation}
All the terms of the summations in $\Gamma_{ij}$ and $C_i$ are proportional to the general terms $\mbox{Re} \left[ e^{i \theta} \bra{\bar{0}} \widetilde{V}_{k,i}^{\dagger} \widetilde{V}_{\ell,j} \ket{\bar{0}} \right]$ and 
$\mbox{Re} \left[ e^{\i \theta} \bra{\bar{0}} \widetilde{V}_{k,i}^{\dagger} h_{\alpha} V \ket{\bar{0}} \right]$ respectively. These quantities can be computed by applying the quantum circuits of Fig. 2. 
In this way, it is possible to compute $\Gamma_{ij}$ and $C_i$ by applying the circuits of Fig. 2 repeatedly. 

After obtaining the ground state $\ket{g} = \ket{\phi_0}$ and corresponding eigenvalue $\lambda_0$, one can compute the first excited state $\ket{\phi_1}$ by changing $A$ as 
\begin{equation}
\label{vqa-12}
A \rightarrow A' = A + \mu B \ket{g} \bra{g} B
\end{equation}
where $\ket{g}$ is normalized as $\bra{g} B \ket{g} = 1$. The parameter $\mu$ is chosen as $\mu > \lambda_1 - \lambda_0$. Since we do not know $\lambda_1$, we should choose $\mu$ sufficiently large. 
Repeating this procedure one can compute the full spectrum of the GEE (\ref{gevalue-1}). 

\section{Numerical Example I: Case for regular $B$}

\begin{figure}[ht!]
\begin{center}
\includegraphics[height=5.0cm]{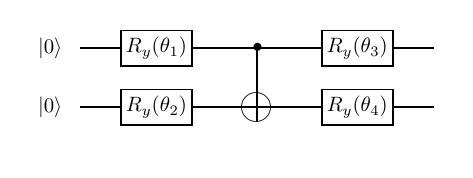} 
\includegraphics[height=3.0cm]{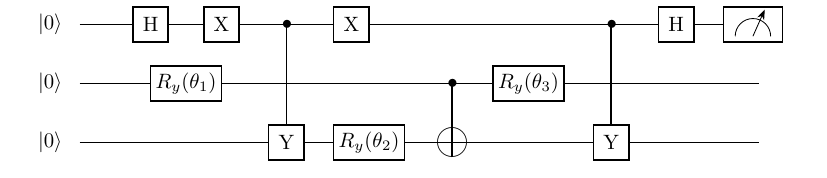}

\caption[fig3]{(Color online) (a) The four-parameter dependent quantum state for numerical example.  (b) Quantum circuit for $\Gamma_{24}$. 
 }
\end{center}
\end{figure}

In this section we apply the Euclidean time method introduced in the previous section to Eq. (\ref{gevalue-1}), where 
\begin{eqnarray}
\label{toy-1}
&&A = I_2 \otimes I_2 + 0.4 Z \otimes I_2 + 0.4 I_2 \otimes Z + 0.2 X \otimes X               \\    \nonumber
&&B = I_2 \otimes I_2 + 0.3 Z \otimes I_2 + 0.4 I_2 \otimes Z + 0.2 Z \otimes Z.
\end{eqnarray}
In this case $A - F B = \sum_{\alpha = 0}^4 \Lambda_{\alpha} h_{\alpha}$, where 
\begin{eqnarray}
\label{toy-2}
&&\Lambda_0 = 1 - F  \hspace{.4cm} \Lambda_1 = 0.4 - 0.3 F  \hspace{.4cm}  \Lambda_2 = 0.4 (1 - F)  \hspace{.4cm}  \Lambda_3 = 0.2  \hspace{.4cm}  \Lambda = -0.2 F   \\   \nonumber
&& h_0 = I_2 \otimes I_2  \hspace{.4cm} h_1 = Z \otimes I_2   \hspace{.4cm}  h_2 = I_2 \otimes Z  \hspace{.4cm}  h_3 = X \otimes X  \hspace{.4cm}  h_4 = Z \otimes Z.
\end{eqnarray}
We choose the state $\ket{\psi (\tau)}$ as four-parameter state shown in Fig. 3(a). Then, it is straightforward to construct the quantum circuits for $\Gamma_{ij}$ and $C_i$. 
For example, the quantum circuit for $\Gamma_{24}$ is plotted in Fig. 3(b).

\begin{figure}[ht!]
\begin{center}
\includegraphics[height=5.0cm]{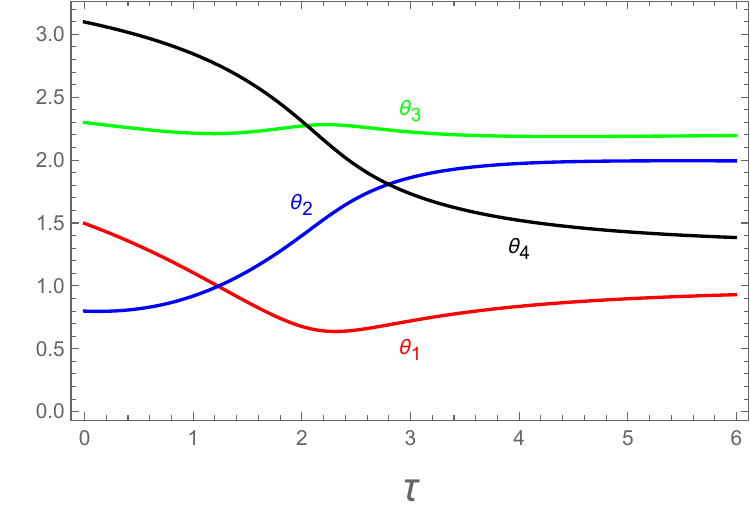} 
\includegraphics[height=5.0cm]{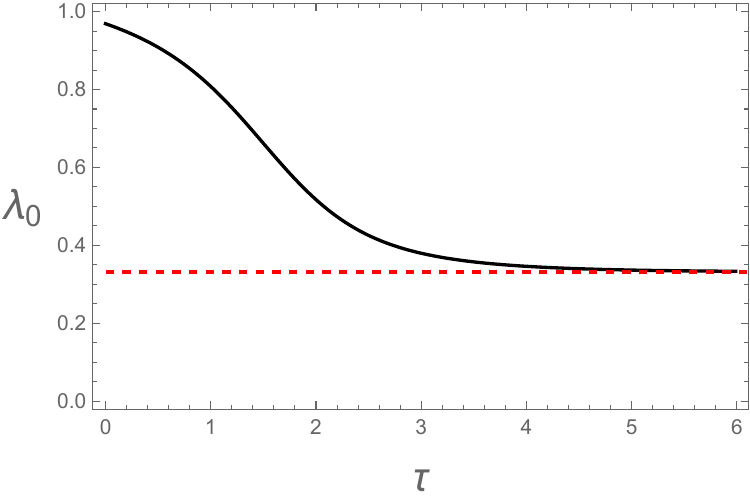}

\caption[fig4]{(Color online)  (a) The Euclidean time evolution of the parameters $\theta_i \hspace{.2cm} (i = 1, 2, 3, 4)$. The initial values of the parameters are chosen as  $\theta_1 = 1.5$, $\theta_2 = 0.8$, $\theta_3 = 2.3$, and $\theta_4 = 3.1$.
(b) The imaginary time evolution of $F(\tau)$. We choose $\delta \tau = 0.01$. It approaches to $0.333262$ as increasing $\tau$. The dashed line corresponds 
to the exact value.  }
\end{center}
\end{figure}

The Euclidean time evolution of the parameters $\theta_i$ and  $\lambda_0$ are plotted in Fig. 4(a) and Fig. 4(b) respectively. The lowest eigenvalue $\lambda_0$ approaches to $0.333262$ when $\tau$ 
approaches to $6$. In this limit the parameters approach 
\begin{equation}
\label{toy-3}
\theta_1 = 0.92987  \hspace{1.0cm}  \theta_2 = 1.99389  \hspace{1.0cm} \theta_3 = 2.19508  \hspace{1.0cm}  \theta_4 = 1.38469.
\end{equation}
The corresponding eigenstate is 
\begin{equation}
\label{toy4}
\ket{\phi_0} = 0.228442 \ket{00} + 0.044591 \ket{01} - 0.032439 \ket{10} - 1.340530 \ket{11}
\end{equation}
where Eq. (\ref{gevalue-2}) is used for normalization.This is very close to the exact eigenstate
\begin{equation}
\label{toy5}
\ket{\psi_0} = 0.229362 \ket{00} - 1.34168 \ket{11}.
\end{equation}
In order to examine how much $\ket{\phi_0}$ is close to $\ket{\psi_0}$, one can compute the fidelity, which results in  $|\bra{\phi_0} \psi_0 \rangle|^2 = 0.998358$, where the usual normalization is used.

\begin{figure}[ht!]
\begin{center}
\includegraphics[height=5.0cm]{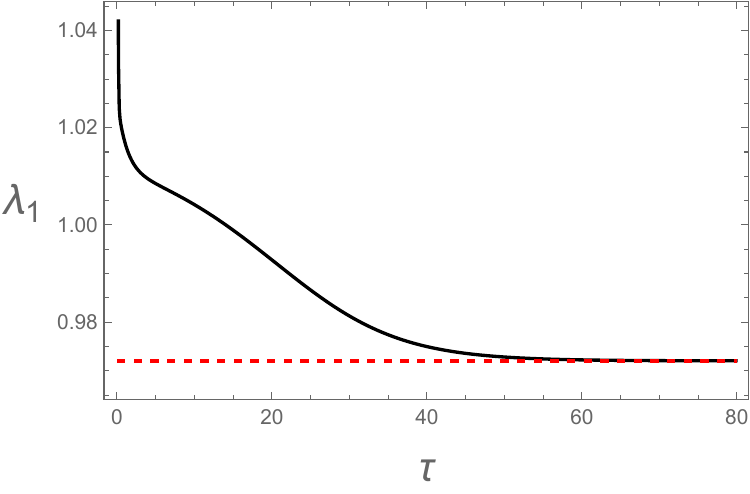} 
\includegraphics[height=5.0cm]{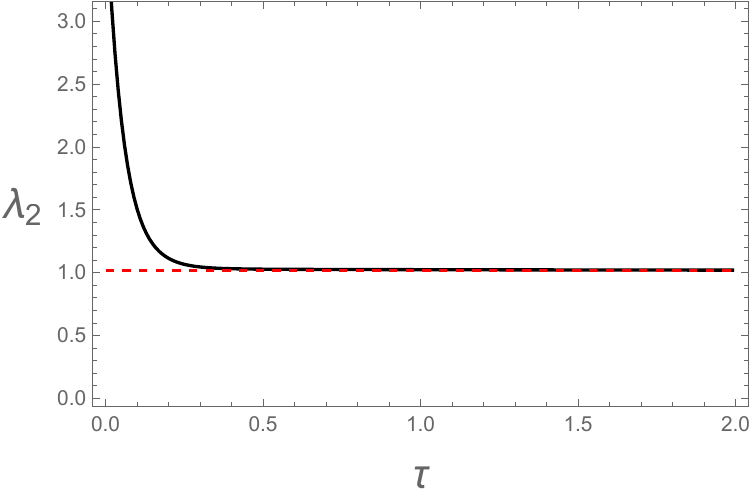}
\includegraphics[height=5.0cm]{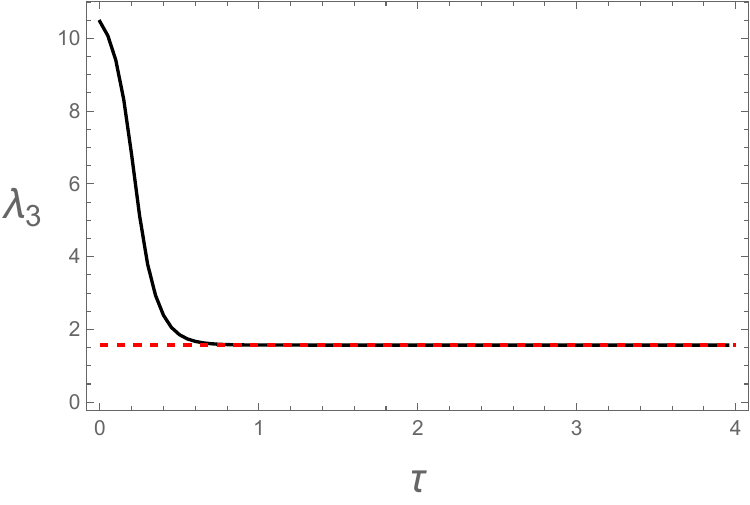}

\caption[fig5]{(Color online) (a) The Euclidean time evolution of $\lambda_1$. We choose $\delta \tau = 0.1$ and $\mu= 10$. It approaches to $0.972050$ as increasing $\tau$. 
(b) The Euclidean time evolution of $\lambda_2$. We choose $\delta \tau = 0.01$ and $\mu_1 = \mu_2 = 10$. It approaches to $1.02106$ as increasing $\tau$.
(c) The Euclidean time evolution of $\lambda_3$. We choose $\delta \tau = 0.05$. It approaches to $1.56964$ as increasing $\tau$.
The dashed lines in (a), (b), and (c) correspond to the exact values.}
\end{center}
\end{figure}

\begin{center}
\begin{tabular}{c|c|c|c|c} \hline \hline
               &  $\lambda_0$   &   $\lambda_1$   &     $\lambda_2$   &   $\lambda_3$                                \\    \hline
imaginary time method  &  \hspace{.1cm}    $0.33326$ \hspace{.1cm}  &  \hspace{.1cm}  $0.97205$ \hspace{.1cm}  & \hspace{.1cm}  $1.02106$ \hspace{.1cm}  & \hspace{.1cm}  $1.56964$                                      \\  \hline 
exact values  &  $0.33162$  &   $0.97204$  & $1.01575$  &  $1.56765$   \\  \hline  \hline
\end{tabular}

\vspace{0.2cm}
Table I:Comparison of result of  the Euclidean time method with exact values.
\end{center}

In order to compute the first-excited eigenvalue we should change the matrix $A$ as $A' = A + \mu B \ket{\phi_0} \bra{\phi_0} B$, where 
\begin{eqnarray}
\label{toy6}
&& B \ket{\phi_0}\bra{\phi_0} B = 0.1599 I_2 \otimes I_2 -0.1459 X \otimes X + 0.1450 Y \otimes Y + 0.1590 Z \otimes Z     \\    \nonumber
&& \hspace{2.4cm} -0.0652 I_2 \otimes Z - 0.0652 Z \otimes I_2 + 0.0166 I_2 \otimes X - 0.0168 X \otimes I_2                                                 \\    \nonumber
&&\hspace{2.4cm}+ 0.0041 X \otimes Z - 0.0030 Z \otimes X.
\end{eqnarray}
Then, we should modify the quantum circuits for $F(\tau)$ and $C_i (\tau)$ to include Eq. (\ref{toy6}). In Fig. 5a the Euclidean time evolution of $\lambda_1$ is plotted, where 
$\delta \tau = 0.1$ and $\mu = 10$ are chosen. This figure shows that $\lambda_1$ approaches to $0.97205$ when $\tau$ approaches to $80$. The second-excited eigenvalue can be 
computed by changing $A$ as $A'' = A + \mu_1 B \ket{\phi_0} \bra{\phi_0} B + \mu_2 B \ket{\phi_1} \bra{\phi_1} B$. The Euclidean time evolution of $\lambda_2$ is plotted in 
Fig. 5b, where $\delta \tau = 0.01$ and $\mu_1 = \mu_2 = 10$ are chosen. The eigenvalue $\lambda_2$ approaches to $1.02106$ when $\tau$ approaches to $2.0$. 
Similarly, the Euclidean time evolution of $\lambda_3$ is plotted in Fig. 5c. The eigenvalue $\lambda_3$ approaches to $1.56964$ at the large $\tau$. The eigenvalues computed by the Euclidean 
time method are compared with the exact values in Table I. Table I shows that the eigenvalues computed by the Euclidean time method coincides with the exact values within $99.5 \%$. 

The satisfactory accuracy of our results is mainly due to the fact that we use the qiskit (version $0.36.2$) in classical computer. If, however, we use the real quantum computer, the discrepancy between numerical and exact results would be increased due to 
the noise effect. 
For this case we have to use the noise mitigation process appropriately. 
Few years ago, a quantum algorithm was proposed to exactly and efficiently discuss the effect of noise on the system\cite{wang2018} in the photosynthetic energy transfer.

\section{Numerical Example II: Case for singular $B$}

\begin{figure}[ht!]
\begin{center}
\includegraphics[height=5.0cm]{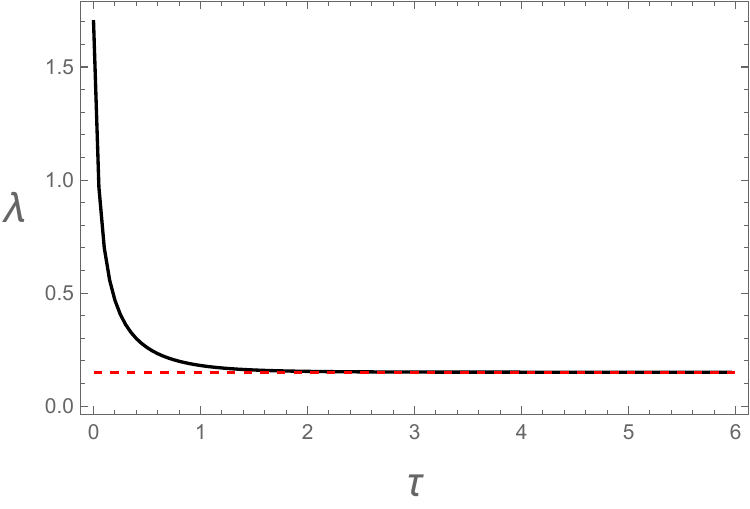} 
\includegraphics[height=5.0cm]{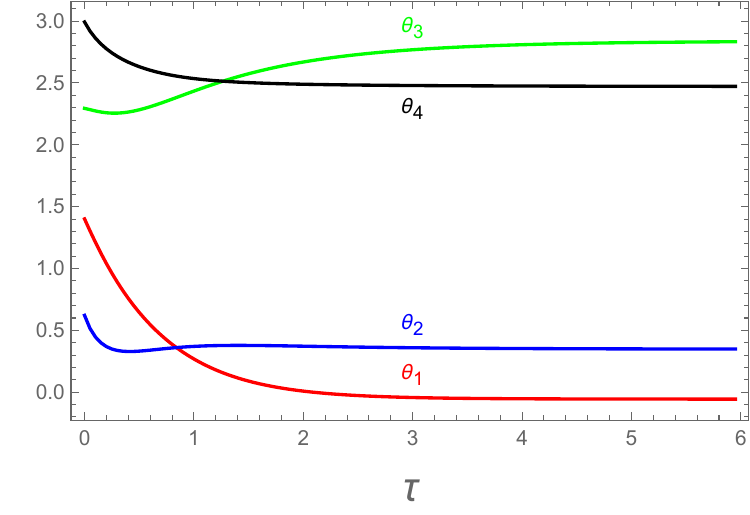}

\caption[fig6]{(Color online) (a) The Euclidean time evolution of $\lambda$. We choose $\delta \tau = 0.05$. As expected it approaches to $0.150005$ as increasing $\tau$.
(b) The Euclidean time evolution of the parameters $\theta_i \hspace{.2cm} (i = 1, 2, 3, 4)$. The final values of the parameters are reduced to $\theta_1 = -0.056$, $\theta_2 = 0.349$, $\theta_3 = 2.833$, and $\theta_4 = 2.473$. }
\end{center}
\end{figure}

In order to confirm that our formalism also can be applied for the singular $B$,  we consider another numerical example when $B$ is singular operator in this section. 
In order to explore this issue, let us choose $A$ and $B$ in the form:
\begin{eqnarray}
\label{singular-1}
A = \left(    \begin{array}{cccc}
              a_1   &  0  &  0  &  b            \\
              0  &  a_2  &  b  &  0             \\
              0  &  b  &  a_3  &  0             \\
              b  &  0  &  0  &  a_4             
                 \end{array}                     \right)    \hspace{1.0cm}
B = \left(    \begin{array}{cccc}
              1  &  1  &  1  &  1              \\
              1  &  1  &  1  &  1              \\
              1  &  1  &  1  &  1              \\
              1  &  1  &  1  &  1               
                  \end{array}               \right).
\end{eqnarray}
In this case the eigenvalue and corresponding eigenvector can be computed analytically. 
Unlike the usual eigenvalue equation one can show that this system generates single eigenvalue in the form
\begin{equation}
\label{s_eigenvalue}
\lambda = \frac{a_1 a_2 a_3 a_4 - (a_1 a_4 + a_2 a_3) b^2 + b^4}{Q}
\end{equation}
where
\begin{equation}
\label{s_eigenvalue-2}
Q = a_1 a_2 a_3 + a_1 a_2 a_4 + a_1 a_3 a_4 + a_2 a_3 a_4 - 2 (a_1 a_4 + a_2 a_3) b - (a_1 + a_2 + a_3 + a_4) b^2 + 4 b^3.
\end{equation}
The corresponding eigenvector can be written as 
\begin{eqnarray}
\label{s_eigenvector}
&&\ket{\Phi} = \frac{1}{Q} \Bigg[ (a_4 - b) (a_2 a_3 - b^2) \ket{00} + (a_3 - b) (a_1 a_4 - b^2) \ket{01}                    \\   \nonumber
&&    \hspace{3.0cm}                    + (a_2 - b) (a_1 a_4 - b^2) \ket{10} + (a_1 - b) (a_2 a_3 - b^2) \ket{11}   \Bigg].
\end{eqnarray} 

As a numerical example we choose 
\begin{eqnarray}
\label{s_toy-1}
&&A = I_2 \otimes I_2 + 0.4 Z \otimes I_2 + 0.4 I_2 \otimes Z + 0.2 X \otimes X               \\    \nonumber
&&B = I_2 \otimes I_2 +  I_2 \otimes X + X \otimes I_2 + X \otimes X.
\end{eqnarray}
Then, Eqs. (\ref{s_eigenvalue}) and (\ref{s_eigenvector}) gives
\begin{equation}
\label{s_exact}
\lambda_{exact} = 0.15  \hspace{1.0cm} \ket{\Phi}_{exact} = 0.125 \ket{01} + 0.125 \ket{10} + 0.75 \ket{11}.
\end{equation}

The Euclidean time evolution of the eigenvalue $\lambda$ is plotted in Fig. 6. As expected it approaches to $0.150005$ as increasing $\tau$.
Using the final values of $\theta_j$, one can derive the corresponding eigenvector, which  is $\ket{\Phi} = 0.127 \ket{01} + 0.124 \ket{10} + 0.749 \ket{11}$. It approximately coincides with $\ket{\Phi}_{exact}$.

\section{Numerical Example III: for $8 \times 8$ matrices of $A$ and $B$ }
\begin{figure}[ht!]
\begin{center}
\includegraphics[height=5.0cm]{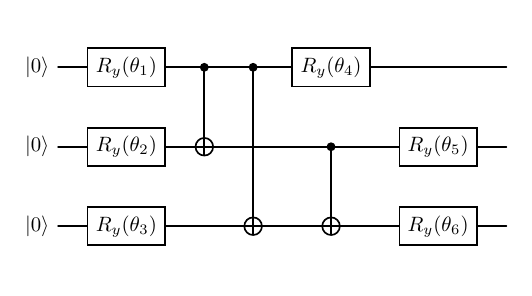}
\includegraphics[height=5.0cm]{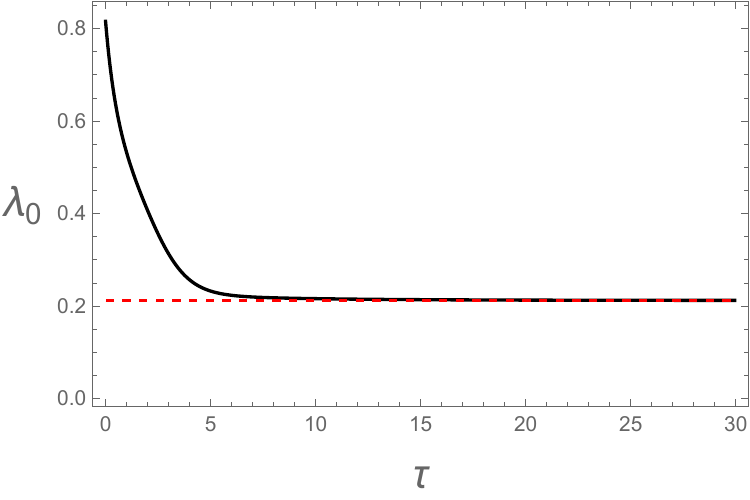} 
\includegraphics[height=5.0cm]{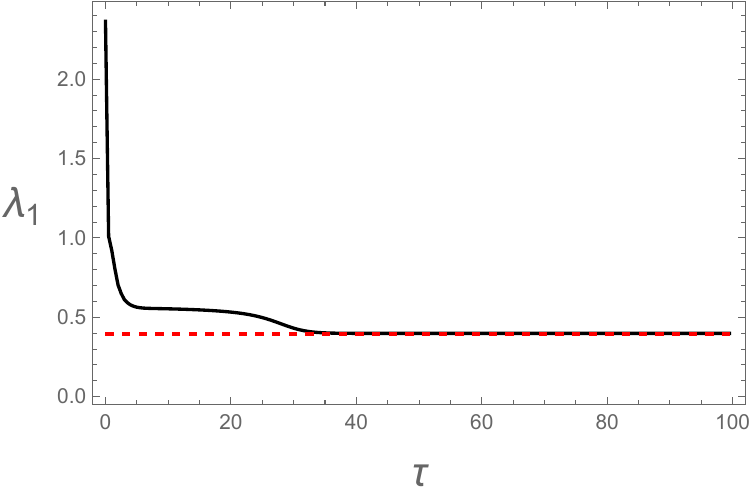}
\caption[fig7]{(a) We chooses $\ket{\psi}$ as six-parameter state. (b) The lowest eigenvalue $\lambda_0$ converges to $0.2126$ when $\tau$ approaches to $30$. The red dashed line corresponds to the exact value $ 0.212465$.     
(c)  The first-excited eigenvalue $\lambda_1$ converges to $0.3988$ when $\tau$ approaches to $100$. The red dashed line corresponds to the exact value $0.394698$.  }
\end{center}
\end{figure}
In this section we apply the Euclidean time method when $A$ and $B$ are $8 \times 8$ matrices as follows:
\begin{eqnarray}
\label{toy-1}
&&A = I_2 \otimes I_2 \otimes I_2 + 0.4 Z \otimes I_2 \otimes X + 0.4 I_2 \otimes Z \otimes X               + 0.2 X \otimes X  \otimes I_2             \\    \nonumber
&&B = I_2 \otimes I_2 \otimes I_2 + 0.3 Z \otimes I_2 \otimes Z + 0.4 I_2 \otimes Z \otimes X + 0.2 Z \otimes Z \otimes X.
\end{eqnarray}
We choose the state $\ket{\psi} (\tau)$ as six-parameter state shown in Fig. 7(a). The lowest eigenvalue $\lambda_0$ approaches to 0.2126 when $\tau$ approaches to $30$, which is shown in Fig. 7(b). 
In order to compute the first-excited eigenvalue $\lambda_1$ we change $A$ as $A'=A + \mu B \ket{\phi_0} \bra{\phi_0} B$, where $\ket{\phi_0}$ is B-orthogonal ground state given by 
$B \ket{\phi_0} \bra{\phi_0} B = \sum_{ijk=0}^3 p_{ijk} \sigma_i \otimes \sigma_j \otimes \sigma_k$ with the nonzero coefficients are 
\begin{eqnarray}
\label{new-1}
&&p_{000} = p_{330} = 0.121   \hspace{.5cm}  p_{001} = p_{033} = p_{303} = p_{331} = 0.085 \hspace{.5cm} p_{003} = p_{333}= -0.080   \\   \nonumber
&&p_{030} = p_{300} = -0.107  \hspace{.5cm} p_{031} = p_{301} = -0.072  \hspace{.5cm} p_{110} = - p_{220} = -0.048                              \\   \nonumber
&&p_{111} = - p_{221} = -0.057  \hspace{.5cm}  p_{113} = - p_{223} = 0.012  \hspace{.5cm} p_{112} = p_{212} = 0.033.
\end{eqnarray}
The coefficient $\mu$ is chosen as $5.0$. Fig. 7(c) shows that $\lambda_1$ approaches 0.3988 when $\tau$ approaches to $100$. In Fig. 7(b) and (c) the red dashed lines correspond to the exact value, which are $ 0.212465$ and $0.394698$, respectively. 
One can compute the higher eigenvalues by similar way. Since this is only tedious repetition, we skip the procedure in this paper.

\section{application to hydrogen atom: Perturbation method}

In this section we examine how to compute the electric polarizability ${\cal P}$ of the hydrogen atom by applying the generalized eigenvalue equation (\ref{gevalue-1}).
If the external electric field is very small, it can be derived by perturbation method, which was studied in Ref. \cite{gee1969}. 
In the following we will review Ref. \cite{gee1969} and in next section same problem is analyzed by applying the Euclidean time method.

Let us consider the Schr\"{o}dinger equation for the hydrogen-like atom with atomic number $Z$.
If we set the energy eigenvalue as $E = - \alpha^2/2$, the Schr\"{o}dinger equation can be converted into the GEE (\ref{gevalue-1}), where
\begin{equation}
\label{hydro-1}
A = \frac{1}{r}, \hspace{1.0cm} B = -\frac{1}{2} {\bm \triangledown}^2 + \frac{1}{2} \alpha^2, \hspace{1.0cm} \lambda_n = \frac{1}{Z} = \frac{1}{n \alpha}.
\end{equation}
In this case the eigenvector $\ket{n, \ell, m}$ should be $B$-normalized, i.e. $\bra{n_1, \ell_1, m_1} B \ket{n_2m \ell_2, m_2} = \delta_{n_1, n_2} \delta_{\ell_1, \ell_2} \delta_{m_1,m_2}$.
Then, it is straightforward to show 
\begin{equation}
\label{hydro-wavef-1}
 \psi_{n,\ell, m} =  \sqrt{\frac{4 \alpha \Gamma(n - \ell)}{n \Gamma(n + \ell + 1)}} e^{-\alpha r} (2 \alpha r)^{\ell} L_{n - \ell - 1}^{2 \ell + 1} (2 \alpha r) Y_{\ell, m} (\theta, \phi)
\end{equation}
where $Y$ and $L$ refer to spherical harmonics and generalized Laguerre polynomials. It is worthwhile noting that the normalization constant is different from the case of usual normalization constant by a factor $\alpha$.

If we apply the external electric field ${\cal E}$ along the $z$-direction, the operator $A$ is changed into 
\begin{equation}
\label{hydro-2}
A = \frac{1}{r} + \frac{\cal E}{Z}  r \cos \theta.
\end{equation}
Since the generalized eigenvalue $\lambda_n$ has only discrete spectrum, one can apply the perturbation more easily than usual perturbation because Hamiltonian has in general both discrete and continuum spectra. 
For the case of the ground state ($n=1$), straight calculation shows $\lambda_1$ in a form:
\begin{equation}
\label{hydro-3}
\lambda_1 = \frac{1}{Z} = \frac{1}{\alpha} + \frac{9}{4 Z^2 \alpha^5} {\cal E}^2 + {\cal O} \left( {\cal E}^3 \right).
\end{equation}
Solving Eq. (\ref{hydro-3}) we can conjecture $\alpha \sim Z \left(1 + \frac{9}{4} \frac{{\cal E}^2}{Z^6}  \right)$, which results in the ground state energy $E_1$ as
\begin{equation}
\label{hydro-g}
E_1 = - \frac{1}{2} \alpha^2 \sim -\frac{Z^2}{2} - \frac{9}{4} \frac{{\cal E}^2}{Z^4}.
\end{equation}
It is interesting to note that the field-dependent term in $E_1$ is proportional to $1/Z^4$.
Thus, the electric polarizability for the hydrogen atom is 
\begin{equation}
\label{hydro-5}
{\cal P} = \frac{1}{\cal E} \frac{d}{d {\cal E}} \left(\frac{9}{4} \frac{{\cal E}^2}{Z^4} \right) \Bigg{|}_{Z=1} = \frac{9}{2}
\end{equation}
in atom units. This was derived by making use of usual perturbation method in Ref.\cite{schiff}.

In order to explore this issue in quantum computer, we need to convert $A$ and $B$ as matrix forms by using mappings to qubit. If this is possible, one can compute the electric polarizability without relying on the perturbation theory. 
However, we do not know how to derive the Jordan-Wigner\cite{jw1,jw2,jw3} or Bravyi-Kitaev\cite{bk1,bk2}
matrix forms of $A$ and $B$. 
In spite of this fact, one can apply the Euclidean time method to the same atomic physics issue by introducing the proper basis in Hilbert space. This is discussed in next section.


\section{application to hydrogen atom: Numerical Method}

\begin{figure}[ht!]
\begin{center}
\includegraphics[height=5.0cm]{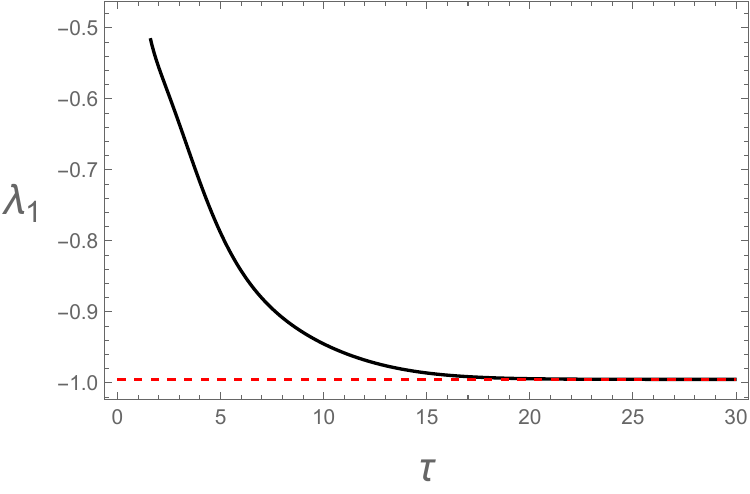}
\includegraphics[height=5.0cm]{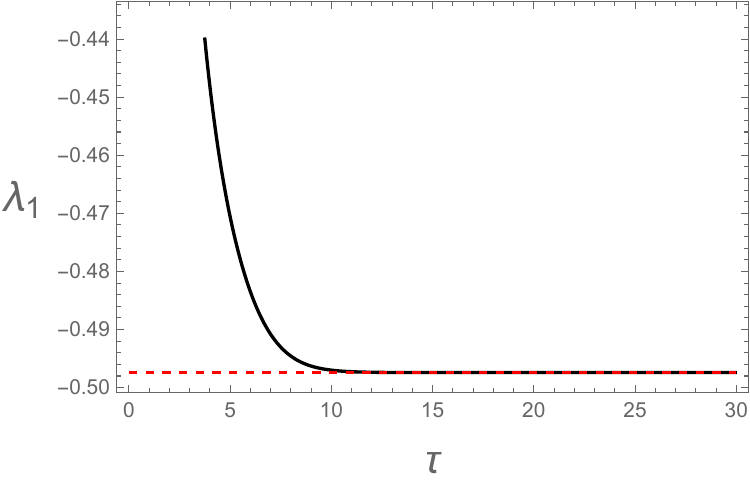} 
\includegraphics[height=5.0cm]{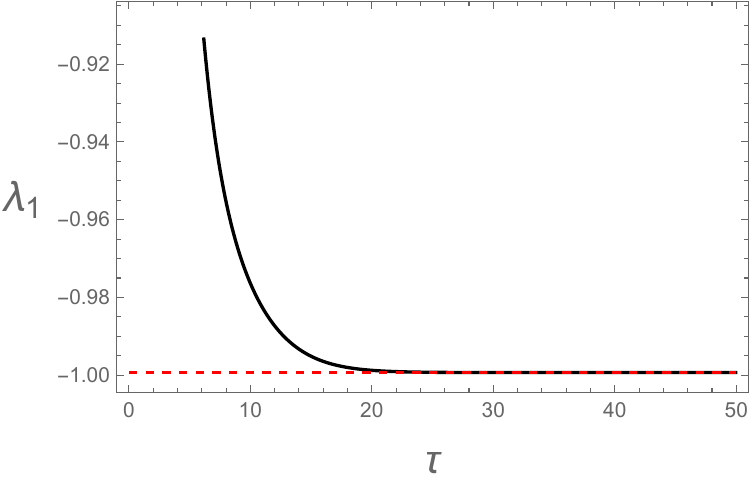}
\includegraphics[height=5.0cm]{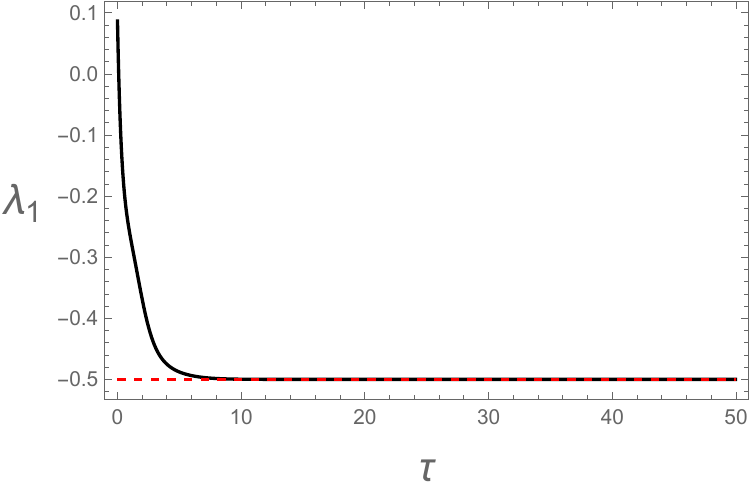}
\caption[fig7]{The Euclidean time evolution of the  lowest eigenvalue $\lambda_1$ for GEE (\ref{gevalue-1}) by choosing $A = \widetilde{A}$ and $B = \widetilde{B}$ 
when (a) $x=0.7$ and $\alpha = -1$, (b) $x=0.7$ and $\alpha = -2$, (c) $x=0.8$ and $\alpha = -1$, (d) $x=0.8$ and $\alpha = -2$.  }
\end{center}
\end{figure}

In this section we would like to apply the Euclidean time method to the atomic physics issue introduced in section VI  without relying on  the Jordan-Wigner or  Bravyi-Kitaev mapping. Instead of the particular mapping to qubit, we will use the matrix representation of $A$ and $B$ 
by introducing the proper basis in Hilbert space. 
Here, let us use the simple nodeless Slater-type orbital (STO) basis\cite{hjo02}
\begin{equation}
\label{sto_basis}
\ket{n, \ell, m} = R_n (r) Y_{\ell,m} (\theta, \phi)
\end{equation}
where
\begin{equation}
\label{sto-1}
R_n (r) = \frac{(2 \xi)^{3/2}}{\sqrt{\Gamma (2 n + 1)}} (2 \xi r)^{n-1} e^{-\xi r}.
\end{equation}
The problem of the STO basis is the fact that it is not completely orthogonal with respect to the principal quantum number as follows:
\begin{equation}
\label{sto-2}
\bra{n', \ell', m'} n, \ell, m \rangle = \frac{\Gamma(n + n' + 1)}{\sqrt{\Gamma(2 n' + 1) \Gamma (2 n + 1)}} \delta_{\ell,\ell'} \delta_{m,m'}.
\end{equation}
Another problem is that this basis involves the free parameter $\xi$. Thus, we have to fix $\xi$ appropriately. Then, the following matrix representations can be derived:
\begin{eqnarray}
\label{matrix_rep}
&&\bra{n', \ell', m'} A \ket{n, \ell, m}   = \frac{\Gamma(n + n')}{\sqrt{\Gamma(2 n' + 1) \Gamma (2 n + 1)}}                     \\      \nonumber
&& \times \Bigg[ 2 \xi \delta_{\ell,\ell'} \delta_{m,m'} + \frac{{\cal E}}{Z} \frac{ (n + n' + 1) (n + n')}{2 \xi} \Bigg\{\sqrt{\frac{(\ell - m + 1) (\ell + m + 1)}{(2 \ell + 1) (2 \ell + 3)}} \delta_{\ell',\ell+1} \delta_{m,m'}                   \\     \nonumber
&& \hspace{8.0cm} + \sqrt{\frac{(\ell - m ) (\ell + m )}{(2 \ell -1) (2 \ell + 1)}} \delta_{\ell',\ell-1} \delta_{m,m'} \Bigg\} \Bigg]                                                \\    \nonumber
&&\bra{n', \ell', m'} B \ket{n, \ell, m}                 \\    \nonumber
&&= \frac{\Gamma(n + n'-1)}{2 \sqrt{\Gamma(2 n' + 1) \Gamma (2 n + 1)}} \Bigg[ \xi^2 \left\{ 4 \ell (\ell + 1) + (n + n') - (n - n')^2 \right\}                                    \\    \nonumber
&&\hspace{8.0cm} + \alpha^2 (n + n') (n + n'-1) \Bigg]  \delta_{\ell,\ell'} \delta_{m,m'}.
\end{eqnarray}
In principle, the matrix representations of $A$ and $B$ are $\infty \times \infty$ dimensional. For the numerical calculation, therefore, we need to truncate them. For example, if we truncate $n, n' \geq 3$, we have $5 \times 5$ matrices of $A$ and $B$ as follows:
\begin{eqnarray}
\label{truncate-1}
&&A = \left(     \begin{array}{ccccc}
x \alpha  &  \frac{x \alpha}{\sqrt{3}}  & 0  &  \frac{{\cal E}}{x Z \alpha}  &    0       \\
\frac{x \alpha}{\sqrt{3}}  &   \frac{x \alpha}{2}  &  0  &  \frac{5 {\cal E}}{2 \sqrt{3} x Z \alpha}  &  0    \\
0  &  0  &  \frac{x \alpha}{2}  &  0  &  0                                                                        \\
 \frac{{\cal E}}{x Z \alpha}  &  \frac{5 {\cal E}}{2 \sqrt{3} x Z \alpha}  &  0  &  \frac{x \alpha}{2}  &  0   \\  
 0  &  0  &  0  &  0  &  \frac{x \alpha}{2}
                  \end{array}      \right)                                                                                                 \\   \nonumber
&&B = \left(      \begin{array}{ccccc}
\frac{(1 + x^2) \alpha^2}{2}  &  \frac{(3 + x^2) \alpha^2}{4 \sqrt{3}}  &  0  &  0  &  0        \\
\frac{(3 + x^2) \alpha^2}{4 \sqrt{3}}  &  \frac{(3 + x^2) \alpha^2}{6}  &  0  &  0  &  0        \\
0  &  0  &  \frac{(1 + x^2) \alpha^2}{2}  &  0  &  0                                                              \\
0  &  0  &  0  &  \frac{(1 + x^2) \alpha^2}{2}  &  0                                                              \\
0  &  0  &  0  &  0  &  \frac{(1 + x^2) \alpha^2}{2}
                  \end{array}    \right)
\end{eqnarray}
where $x$ is defined as $\xi = x \alpha$.
Since the qubit system only needs $2^n \times 2^n$ matrix, we change the $5 \times 5$ matrices into $\widetilde{A} = \left( \begin{array}{cc} A  &  0  \\   0  & I_3  \end{array} \right)$ and $\widetilde{B} = \left( \begin{array}{cc} B  &  0  \\   0  & I_3  \end{array} \right)$, where 
$I_3$ is a $3 \times 3$ identity matrix.  

Now, $\widetilde{A}$ and $\widetilde{B}$ are $8 \times 8$ matrices with free parameters $x$, $\alpha$, $Z$, and ${\cal E}$. 
With aid of Mathematica 13.1 one can show that when ${\cal E} \ll 0$,  the lowest eigenvalue $\lambda_1$ of the GEE (\ref{gevalue-1}) with  $\widetilde{A}$ and $\widetilde{B}$ is similar to Eq. (\ref{hydro-3}) in a form 
\begin{equation}
\label{sto-3}
\lambda_1 = \frac{1}{Z} = g_1(x) \frac{1}{\alpha} + g_2(x) \frac{{\cal E}^2}{Z^2 \alpha^5} +  {\cal O} \left( {\cal E}^3 \right)
\end{equation}
where $g_1$ and $g_2$ depend only on $x$. Then, the electric polarizability becomes ${\cal P} (x) = \frac{2 g_2(x)}{g_1(x)^3}$.

 In the following  we will compute $g_1(x)$ and $g_2(x)$ by applying the Euclidean method as follows. We fix ${\cal E} = 0.01$ and $Z=1$ for simplicity. 
 Given $x = x_*$ we compute $\lambda_1$ by the Euclidean time method for two different $\alpha$. Solving two coupled equations of Eq. (\ref{sto-3}) one can compute $g_1(x_*)$
 and $g_2(x_*)$.  For example, Fig. 8(a) and (b) correspond to the Euclidean time evolution of $\lambda_1$ when $(x, \alpha) = (0.7, -1)$ and $(x, \alpha) = (0.7, -2)$  respectively. 
 Fig. 8(c) and (d) correspond to $x=0.8$ with same values of $\alpha$. We use the same initial state of section V given in Fig. 7(a). Our numerical result is summarized in Appendix A as a Table II.
 From Table II $2 g_2 (x) / g_1(x)^3$ is maximized at $x_* = 0.9$ and at this point we have ${\cal P} (x_*) = 4.2665$. This is slightly less than the perturbation result $4.5$. 
 Of course, different truncation yields different matrix representations of $A$ and $B$. If we truncate $n, n' \geq 101$, the dimension of $\widetilde{A}$ and $\widetilde{B}$ becomes $2^{19} \times 2^{19}$. 
In this case we need  at least $20$-qubit  quantum computer for the computation of the electric polarizability using the Euclidean time method. 



\section{Conclusion}

In this paper we apply the Euclidean time method of VQE to the GEE (\ref{gevalue-1}). For the purpose of this we slightly modified the usual  imaginary time method of VQE presented in Ref.\cite{euclidean-1}.
We applied our formalism to the three numerical examples. It is shown that 
the Euclidean time technique introduced in this paper works very well for all example. 
Finally, we apply our method to the hydrogen atom system and compute the electric polarizability when the external electric field is ${\cal E} = 0.01$. 
It turns out that the polarizability is $4.2665$, which is slightly less than the perturbation result $4.5$.

There are lot a issues we need to address. How to compute the electric polarizability of the hydrogen atom when ${\cal E}$ is large? 
In this case the perturbation method is useless. How to extend our method to other atoms such as helium or lithium?
It is of interest to apply our formalism to the real physical,  chemical, and fluid problems.



{\bf Acknowledgement}:
This work was supported by the National Research Foundation of Korea(NRF) grant funded by the Korea government(MSIT) (No. 2021R1A2C1094580).

\vspace{1.0cm}

{\bf Data Availability}:
 The datasets generated during and/or analyzed during the current study are available from the corresponding author on reasonable request.
 
 \vspace{1.0cm}

{\bf Conflict of Interest}: 
The authors declare that they have no known competing financial interests or personal relationships that could have appeared to influence the work reported in this paper.

\end{document}